\documentclass[useAMS]{mn2e}
\usepackage{graphicx}

 \usepackage{url}
\usepackage{times}
\def\spose#1{\hbox to 0pt{#1\hss}}
\newcommand\lsim{\mathrel{\spose{\lower 3pt\hbox{$\mathchar"218$}}
     \raise 2.0pt\hbox{$\mathchar"13C$}}}
\newcommand\gsim{\mathrel{\spose{\lower 3pt\hbox{$\mathchar"218$}}
     \raise 2.0pt\hbox{$\mathchar"13E$}}}
\def\sc{Schwarzschild}
\def\ltsima{$\; \buildrel < \over \sim \;$}
\def\lsim{\lower.5ex\hbox{\ltsima}}
\def\gtsima{$\; \buildrel > \over \sim \;$}
\def\gsim{\lower.5ex\hbox{\gtsima}}

\def\ergcms{{\rm\thinspace erg \thinspace cm^{-2} \thinspace s^{-1}}}
\def\ergs{{\rm\thinspace erg \thinspace s^{-1}}}

\title[Jet--accretion relation in blazars, observed by {\it NuSTAR}]
{Extremes of the jet--accretion power relation of blazars, as explored by {\it NuSTAR}
	}
\author[Sbarrato et al.]{T.\  Sbarrato$^1$\thanks{E--mail: tullia.sbarrato@unimib.it}, 
G.\ Ghisellini$^{2}$, G.\ Tagliaferri$^{2}$, M.\ Perri$^{3,4}$,  
G.\ M.\ Madejski$^{5}$, 
D.\ Stern$^{6}$, 
\newauthor{
S.\ E.\ Boggs$^{7}$, F.\ E.\ Christensen$^{8}$, 
W.\ W.\ Craig$^{8,9}$, 
C.\ J.\ Hailey$^{10}$, F.\ A.\ Harrison$^{11}$, }
\newauthor{W.\ W.\ Zhang$^{12}$} \\ \\ 
$^1$Dipartimento di Fisica ``G.\ Occhialini", 
			Universit\`a di Milano -- Bicocca, piazza della Scienza 3, I--20126 Milano, Italy \\
$^2$INAF -- Osservatorio Astronomico di Brera, via E. Bianchi 46, I--23807 Merate, Italy \\
$^3$ASI Science Data Center, via del Politecnico, I--00133 Roma, Italy \\
$^4$INAF -- Osservatorio Astronomico di Roma, via Frascati 33, 
		      	I--00040 Monteporzio Catone, Italy \\
$^5$Kavli Institute for Particle Astrophysics and Cosmology,
			SLAC National Accelerator Laboratory, Menlo Park, CA 94025, USA \\
$^6$Jet Propulsion Laboratory, California Institute of Technology, Pasadena, 
			CA 91109, USA \\
$^7$Space Sciences Laboratory, University of California, Berkeley, CA 94720, USA \\
$^8$DTU Space - National Space Institute, Technical University of
			Denmark, Elektrovej 327, 2800 Lyngby, Denmark \\
$^9$Lawrence Livermore National Laboratory, Livermore, CA 94550, USA \\
$^{10}$Columbia Astrophysics Laboratory, Columbia University, New York, 
			NY 10027, USA \\
$^{11}$Cahill Center for Astronomy and Astrophysics, California Institute
			of Technology, Pasadena, CA 91125, USA \\
$^{12}$NASA Goddard Space Flight Center, Greenbelt, MD 20771, USA \\
}

\begin{document}

\pagerange{\pageref{firstpage}--\pageref{lastpage}} \pubyear{2016}

\maketitle
\label{firstpage}

\begin{abstract}

Hard X--ray observations are crucial to study the non--thermal jet emission 
from high--redshift, powerful blazars. 
We observed two bright $z>2$ flat spectrum radio quasars (FSRQs) in hard
X--rays to explore the details of their relativistic jets and their possible variability. 
S5 0014+81 (at $z=3.366$) and B0222+185 (at $z=2.690$) have been observed twice by 
the {\it Nuclear Spectroscopic Telescope Array} ({\it NuSTAR}) simultaneously with 
{\it Swift}/XRT, showing different variability behaviours. 
We found that {\it NuSTAR} is instrumental to explore the variability of powerful high--redshift 
blazars, even when no $\gamma$--ray emission is detected. 
The two sources have proven to have respectively the most luminous accretion disk and the most 
powerful jet among known blazars. 
Thanks to these properties, they are located at the extreme end of 
the jet--accretion disk relation previously found for $\gamma$--ray detected blazars, 
to which they are consistent. 

\end{abstract}



\section{Introduction}
\label{sec-intro}

Blazars are active galactic nuclei (AGN) with their broad--band 
emission dominated by the relativistic jet, oriented close to our 
line of sight. 
The two humps that characterise their spectral energy distribution (SED) 
are the signature of this relativistically beamed emission. 
They are attributed to synchrotron (at low frequencies) and 
inverse Compton (at high frequencies) processes, and 
in the radio/sub--millimeter and X-- /$\gamma$--ray, respectively. 
The electron population involved in the inverse Compton emission 
is thought to interact either with the synchrotron photons involved 
in the low--frequency emission, or with photons coming from 
structures external to the relativistic jet (synchrotron self--Compton, SSC, 
or external Compton, EC, emissions, respectively). 
The latter is likely the primary process in sources that present a 
pronounced dominance of the higher frequencies hump over the synchrotron one. 
This usually happens in the most powerful blazars, i.e.\ 
the flat--spectrum radio quasars (FSRQs). 
These sources are thought to have more sources of seed photons 
for an EC process (i.e. accretion disk, broad line region, torus), 
compared to the BL Lacertae objects (BL Lacs) that have weak
or absent broad lines and no accretion or torus emission
(see e.g. Chiaberge, Capetti \& Celotti 1999; 
Ghisellini et al.  2011, Sbarrato et al.  2012). 

The most immediate signature of the blazar nature of an AGN is its 
emission in the $\gamma$--rays. 
The high--energy hump in very powerful blazar SED, in particular, usually peaks in the MeV--GeV range, 
and therefore it can be easily observed with $\gamma$--ray 
telescopes, such as the Large Area Telescope (LAT) onboard the 
{\it Fermi Gamma--Ray Space Telescope} (Atwood et al.\ 2009). 
The {\it Fermi}/LAT team built an all--sky $\gamma$--ray catalog, 
providing a clear classification of all the sources included in the 
survey, through multifrequency studies.
This provides a complete, all--sky blazar catalog (Ackermann et al.\ 2015).  
Nevertheless, at higher redshifts, {\it Fermi}/LAT is less efficient 
in detecting blazars, even those with a very large bolometric
luminosity.
This is because the most powerful blazars have their high energy peak
at $\sim$MeV energies or below, and this peak is seen redshifted.
This is the reason why the fraction of high redshift blazars (i.e. at $z>2$)
detected in the hard X--rays by the Burst Alert Telescope (BAT) onboard
the {\it Swift} satellite is much larger than for {\it Fermi}/LAT
(see Ajello et al. 2009, Ghisellini et al 2010a).

Indeed, blazars observed so far show a trend: the humps in the SEDs of the 
more powerful ones peak at lower frequencies as compared to less powerful blazars. 
This trend is known as the ``blazar sequence'' (Fossati et al.\ 1998). 
Although the original concept of "blazar sequence" finds confirmation
through  {\it Fermi} blazar data (i.e. Ajello et al.\ 2015; 
Ackermann et al.\ 2015), there is some dispute about its reality
(see e.g. the reviews by Padovani 2007 and Ghisellini et al. 2008).
For instance, Giommi et al. (2012) proposed what they called
a ``simplified blazar scenario", in which they postulate that the shape
of the SED of blazars is uncorrelated with their luminosity.
Then they assume a given probability for the different blazar shape:
there are more blazar with low--frequency peaks than high--frequency ones.
Taking into account the observation constraints and the limiting fluxes
of the current blazar surveys, they can reproduce what is observed.
This should be taken as a test that both the ``blazar sequence" and
the ``simplified scenario" pass, not as a proof that the blazar sequence is wrong.
Both frameworks can describe the considered existing data.
On the other hand, the blazar sequence found an easy physical
explanation in terms of radiative cooling (Ghisellini et al. 1998), while
the simplified scenario is based on the assumed SED distribution,
that has no physical explanation (yet).
Whether the blazar sequence is intrinsic or only a selection effect, 
the most powerful and distant blazars are hardly detected
by {\it Fermi}/LAT, their inverse Compton emission peak being shifted 
towards the (observed) MeV band (Ghisellini et al.\ 2010). 
Hard X--ray instruments, instead, like {\it Swift}/BAT and now the 
{\it Nuclear Spectroscopic Telescope Array} ({\it NuSTAR}; Harrison et al.\ 2013),
are the most suitable instruments now available to investigate 
jet emission in the most powerful blazars at $z\sim2-3$. 

In this paper we report on observations of S5 0014+81 
($00^\circ17^\prime08.5^{\prime\prime}{\rm+81d\,35m\,08s}$, $z=3.366$) 
and B0222+185 ($02^\circ25^\prime04.7^{\prime\prime} {\rm+18d\,46m\,49s}$, $z=2.690$)
by {\it NuSTAR}.
These two blazars have been previously detected in the 3--year all--sky survey
of {\it Swift}/BAT (Ajello et al. 2009, see also Ajello et al. 2012 and Baumgartner et al. 2013),
and are amongst the most powerful blazars ever observed. 
As with other powerful and high--redshift FSRQs, their optical flux shows  
contributions due to thermal emission from the accretion disk, particularly
prominent in S5 0014+813, whose luminosity reaches $\sim 10^{48}$ erg s$^{-1}$
(Ghisellini et al. 2010a).
For both sources, the {\it Swift}/BAT spectrum together with the {\it Fermi}/LAT upper limit 
already constrained the location of the high energy peak, but with a relatively
large uncertainty given the poor spectral slope determination of {\it Swift}/BAT.
This motivated the {\it NuSTAR} observations.

In this work, we adopt a flat cosmology with $H_0=68$ km s$^{-1}$ Mpc$^{-1}$ and
$\Omega_{\rm M}=0.3$, as found by Planck Collaboration XIII (2015). 

\begin{table*} 
\centering
\footnotesize
S5 0014+81 \\
\vskip 0.1 cm
\begin{tabular}{lllllllllll}
\hline
\hline
Date & $\Gamma_1$ & $\Gamma_2$ & $E_{\rm break}$ & 
   $F_{\rm 0.5-2kev}$ & $F_{\rm 2-10kev}$ & $F_{\rm 10-50kev}$ & $\chi^2$ / dof \\
 & & & keV & $\ergcms$ & $\ergcms$ & $\ergcms$ & \\
\hline   
2014 Dec 1 & $1.18^{+0.22}_{-0.27}$ & $1.72\pm0.05$ & $2.22^{+0.83}_{-0.58}$ &
  $1.37^{+0.04}_{-0.1}\times10^{-12}$ & $3.83_{-0.25}^{+0.15}\times10^{-12}$ 
  & $5.98_{-0.71}^{+0.31}\times10^{-12}$ & 151.8 / 136 \\
2015 Jan 23 & $0.64^{+0.44}_{-0.61}$ & $1.65\pm0.05$ & $1.61^{+0.61}_{-0.34}$ &
  $1.31_{-0.14}^{+0.04}\times10^{-12}$ & $3.62_{-0.44}^{+0.22}\times10^{-12}$ 
  & $6.36_{-0.79}^{+0.28}\times10^{-12}$ & 133.2 / 147 \\
\hline
\hline 
\end{tabular}
\vskip 0.3 cm
B0222+185 \\
\vskip 0.1 cm
\begin{tabular}{lllllllllll}
\hline
\hline
Date & $\Gamma_1$ & $\Gamma_2$ & $E_{\rm break}$ & 
   $F_{\rm 0.5-2kev}$ & $F_{\rm 2-10kev}$ & $F_{\rm 10-50kev}$ & $\chi^2$ / dof \\
 & & & keV & $\ergcms$ & $\ergcms$ & $\ergcms$ & \\
\hline   
2014 Dec 24 & $1.10^{+0.19}_{-0.10}$ & $1.56^{+0.09}_{-0.03}$ & $4.77^{+2.61}_{-0.55}$ &
  $2.62_{-0.12}^{+0.07}\times10^{-12}$ & $1.22_{-0.03}^{+0.01}\times10^{-11}$ 
  & $2.67_{-0.07}^{+0.04}\times10^{-11}$ & 483.6 / 504 \\
2015 Jan 18 & $1.06\pm0.12$ & $1.69^{+0.05}_{-0.04}$ & $4.69^{+0.83}_{-0.56}$ &
  $1.86_{-0.12}^{+0.09}\times10^{-12}$ & $8.29_{-0.18}^{+0.08}\times10^{-12}$ 
  & $1.51_{-0.05}^{+0.03}\times10^{-11}$ & 406.8 / 400 \\
\hline
\hline 
\end{tabular}
\caption{Parameters of the X--ray spectral analysis, from the simultaneous fit of 
{\it NuSTAR} and {\it Swift}/XRT. 
The errors are at 90\% level of confidence for spectral index and break energy, 
at 68\% for the fluxes. 
Fluxes are corrected for galactic absorption. 
}
\label{Xspec}
\end{table*}

\section{Observations and Data Analysis}
\label{sec-data}

We performed simultaneous {\it NuSTAR} and {\it Swift} observations 
in two observing periods for each source. 
This section describes the data analysis performed on these new X--ray 
and optical--UV data. 
Along with the new data sets, we consider archival data 
for the overall SED modeling (see \S \ref{sec-mass} and \ref{sec-sed}). 
Specifically, S5 0014+813 radio and IR data are from Ghisellini et al.\ (2009), 
integrated with new IR photometry from the {\it Wide--field Infrared Explorer} ({\it WISE}\footnote{
Data retrieved from \url{http://irsa.ipac.caltech.edu/}
}; Wright et al.\ 2010). 
In the case of B0222+185, archival data were all retrieved through 
the ASI Science Data Center (ASDC\footnote{
\url{http://tools.asdc.asi.it/}}).
Both sources have not been detected by {\it Fermi}/LAT, but we obtained 
some information from this lack of detection with an ``upper limit" on their 
$\gamma$--ray fluxes. 
The sensitivity limit at 5$\sigma$ estimated on 5 years of observations 
gives a good constrain on the high--energy emission of S5 0014+813 and B0222+185.

\subsection{NuSTAR observations}

The {\it NuSTAR} satellite observed S5 0014+81 on
2014 December 21 (obsID 60001098002) and on 2015 January 23 (obsID
60001098004) for total net exposure times of 31.0 ks and 36.4 ks,
respectively. B0222+185 was observed by {\it NuSTAR} on 2014 December 24
(obsID 60001101002) and on 2015 January 18 (obsID 60001101004). The
total net exposure times were 32.0 ks and 37.4 ks, respectively.

The Focal Plane Module A (FPMA) and Focal Plane Module B (FPMB) data
sets were first processed with the {\it NuSTARDAS} software package
(v.1.4.1) jointly developed by the ASI Science Data Center (ASDC,
Italy) and the California Institute of Technology (Caltech,
USA). Event files were calibrated and cleaned with standard filtering
criteria using the {\it nupipeline} task (version 20150316) of
the {\it NuSTAR} CALDB.

The two sources were well detected in the {\it NuSTAR} 3--79 keV energy band. In
both cases the FPMA and FPMB spectra of the target were extracted from
the cleaned event files using a circle of 20 pixel ($\sim49$ arcsec)
radius, while the background was extracted from nearby circular
regions of 40 pixel radius. The ancillary response files were
generated with the {\it numkarf} task, applying corrections for the PSF
losses, exposure maps and vignetting. All spectra were binned to
ensure a minimum of 30 counts per bin.

\subsection{Swift observations}

The {\it Swift} satellite observed S5 0014+81 on 2014 December 21 (obsID
00080003001) and on 2015 January 23 (obsID 00080003002) while
B0222+185 was observed on 2014 December 24 (obsID 00080243001) and on
2015 January 18 (obsID 00080243002). The total net exposure times were 6.5
ks (December 2014) and 6.6 ks (January 2015) for S5 0014+81 and 4.9 ks
(December 2014) and 5.1 ks (January 2015) for B0222+185.

\subsubsection{XRT observations}

{\it Swift}/XRT (Burrows et al. 2005) observations were carried out using the
Photon Counting (PC) CCD readout mode and in the four observations the
sources were well detected in the 0.3-10 keV XRT energy band. The XRT
data sets were first processed with the XRTDAS software package
(v.3.0.0) developed at the ASI Science Data Center (ASDC) and
distributed by HEASARC within the HEASoft package (v. 6.16). In
particular, event files were calibrated and cleaned with standard
filtering criteria with the {\it xrtpipeline} task using the
calibration files available in the version 20140709 of the {\it Swift}/XRT
CALDB.

The energy spectra were then extracted from the calibrated and cleaned
event files. Events for the source spectral analysis were selected
within a circle of 20 pixel ($\sim47$ arcsec) radius, enclosing about 90\% of
the PSF, while the background was extracted from a nearby circular
region of 80 pixel radius. The ancillary response files were
generated with the {\it xrtmkarf} task, applying corrections for the
PSF losses and CCD defects using the cumulative exposure map. The
source spectra were binned to ensure a minimum of 30 counts per bin.

\subsubsection{UVOT observations}

{\it Swift}/UVOT (Roming et al. 2005) observations were performed with all six
optical and UV lenticular filters (namely $V, U, B, W1, M2, W2$). We
performed aperture photometry for all filters in all the observations
using the standard UVOT software distributed within the HEAsoft
package (version 6.16) and the calibration included in the latest
release of the CALDB. Counts were extracted from apertures of 5 arcsec
radius for all filters and converted to fluxes using the standard zero
points (Poole et al. 2008). The fluxes were then de--reddened using the
appropriate values of $E(B-V)$ taken from Schlegel et al. (1998) and
Schlafly et al. (2011) with $A_\lambda/E(B-V )$ ratios calculated for
UVOT filters using the mean Galactic interstellar extinction curve
from Fitzpatrick (1999). No variability was detected within single
exposures in any filter. The processing results were carefully
verified, checking for possible contaminations from nearby objects within
the source apertures and from objects falling within the background apertures.

\subsection{X--ray spectral analysis}
\label{xray_an}

\begin{figure*}
\vskip -1 true cm 
\hskip -0.5cm
\includegraphics[width=0.5\textwidth]{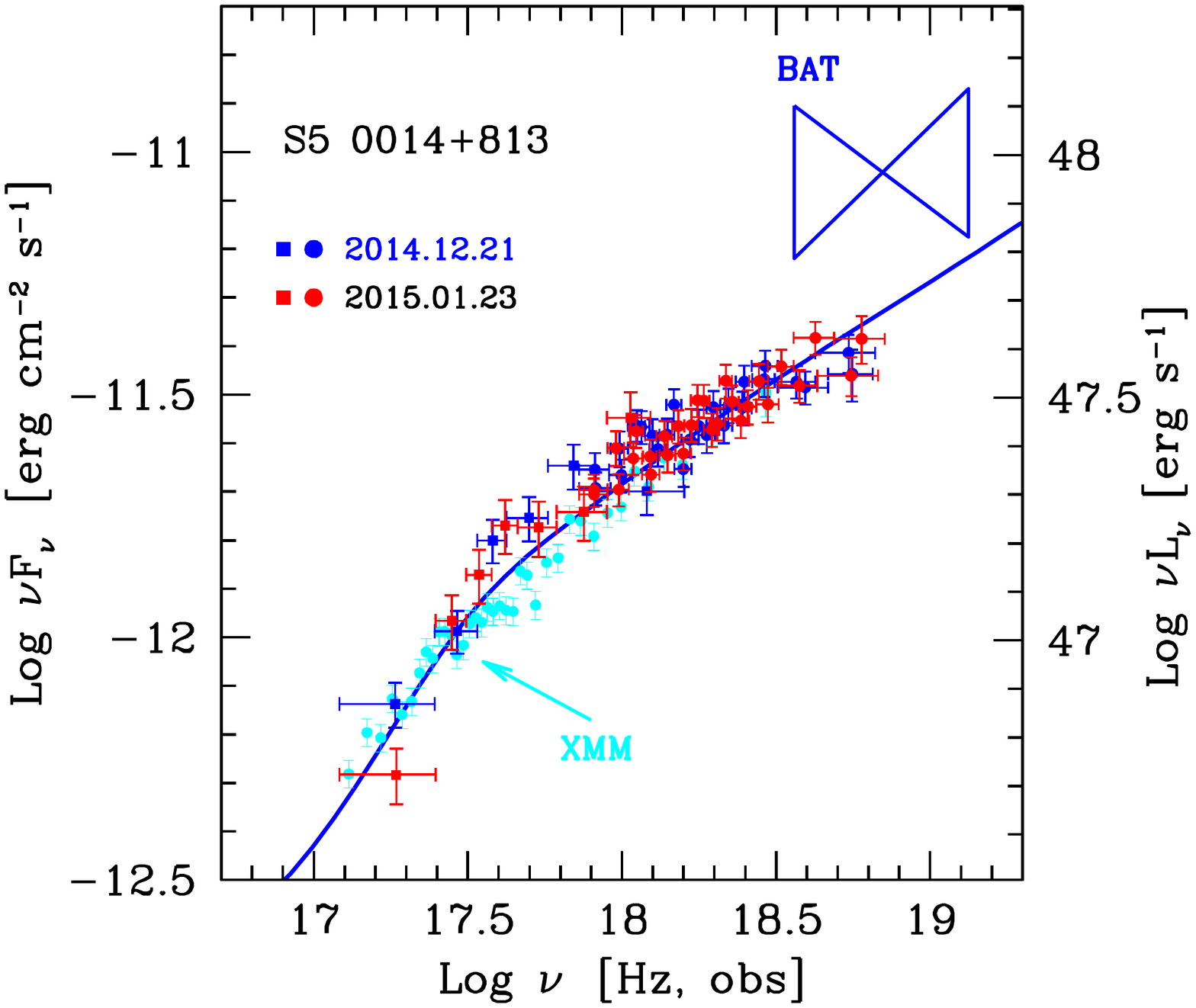}
\includegraphics[width=0.5\textwidth]{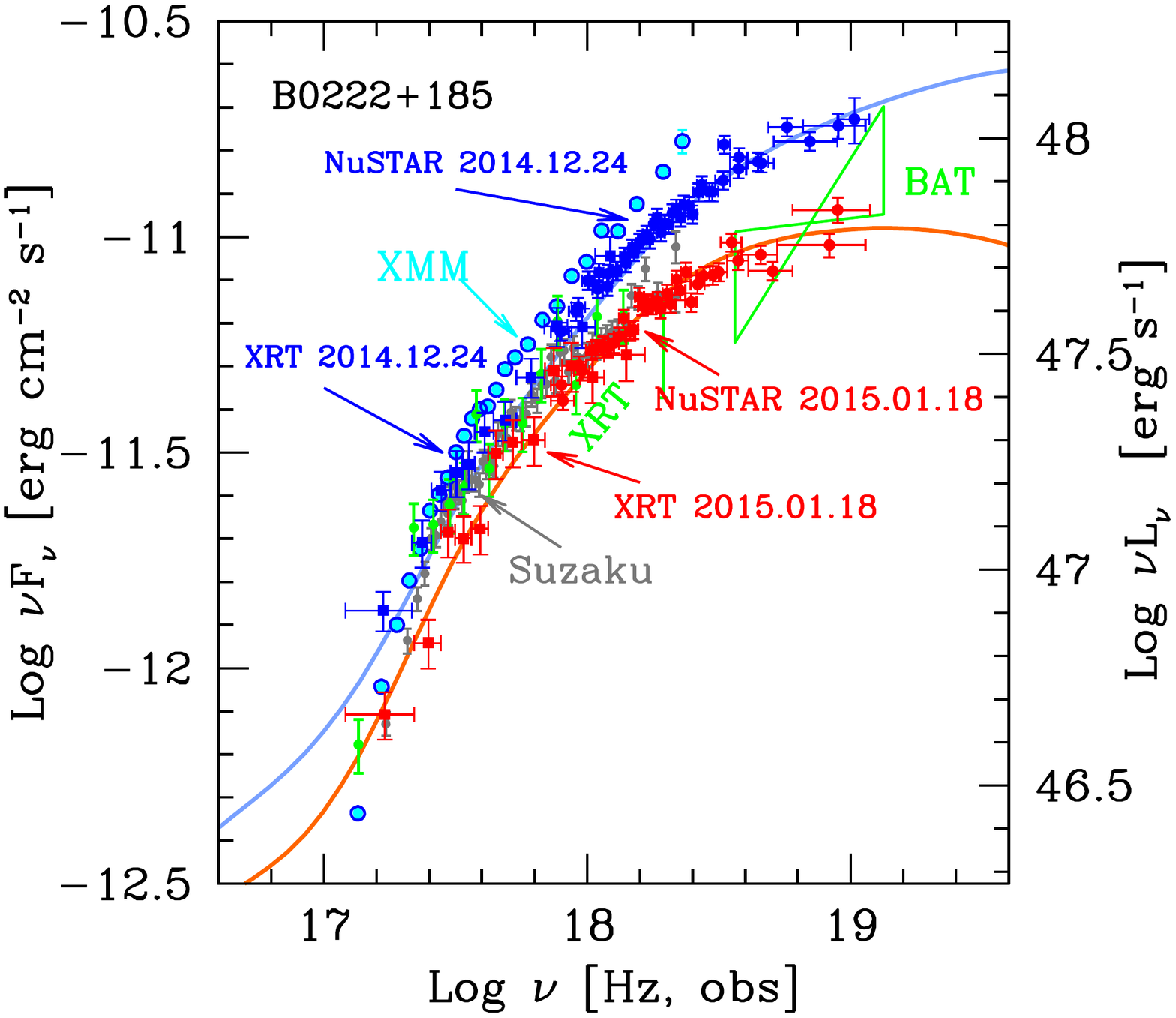}
\vskip -2.5 true cm
\caption{
X--ray spectra of S5 0014+813 and B0222+185, along with the SED models 
described in the text with parameters as in Tables \ref{para} and \ref{powers} 
(shown in solid blue and red lines). 
In both panels, new {\it Swift}/XRT and {\it NuSTAR} data are respectively 
filled squares and circles. The December 2014 observations are in blue, 
the January 2015 ones are in red. 
{\it Left panel}: 
X--ray spectrum of S5 0014+813.
Archival {\it Swift}/BAT data are shown in blue, while  
the cyan circles are archival XMM data, as labelled. 
These data were used and commented in detail in Ghisellini et al.\ (2009; 2010a). 
{\it Right panel}:
X--ray spectrum of B0222+185. 
Archival data are shown in green ({\it Swift}/XRT and BAT), grey ({\it Suzaku}) 
and cyan ({\it XMM--Newton}). 
They were shown and described in detail in Ghisellini et al.\ (2009). 
} 
\label{zoom}
\end{figure*}

The spectral analysis of the December 2014 and January 2015 {\it NuSTAR} and
{\it Swift}/XRT simultaneous observations of S5 0014+81 and B0222+185 were
performed using the XSPEC package. In all four observations a broken
power--law model with an absorption hydrogen--equivalent column density
fixed to the Galactic value of $1.35\times10^{21}$ cm$^{-2}$ (S5 0014+81) and
$9.4\times10^{20}$ cm$^{-2}$ (B0222+185) (Kalberla et al. 2005) was found to
provide a good description of the observed spectra in the 0.3--79 keV
energy band. The inter--calibration factors between the three
instruments ({\it NuSTAR}/FPMA, {\it NuSTAR}/FPMB and {\it Swift}/XRT) were taken into
account adding a multiplicative constant (kept to 1 for
{\it NuSTAR}/FPMA) to the spectral model. We found values in the 2\% range
for {\it NuSTAR}/FPMB and in the 10\% range for {\it Swift}/XRT which are
consistent with the cross--calibration uncertainties for the instruments
(Madsen et al. 2015). The results of the spectral fits
are shown in Table \ref{Xspec}, and Figure \ref{zoom} shows 
the X--ray spectra of the two sources. 

These results describe intrinsic broken power--laws, 
not consistent with absorption. 
We tested different spectral models, namely single power--laws 
with an absorption hydrogen--equivalent column density
fixed to the Galactic value or left free to vary. 
These were possibilities explored by other authors also for these sources 
(e.g.\ Page et al.\ 2005; Piconcelli \& Guainazzi 2005; Tavecchio et al.\ 2007; Eitan \& Behar 2013). 
When we left $N_{\rm H}$ free to vary, the inter--calibration factor between 
{\it Swift}/XRT and {\it NuSTAR}/FPMA--FMPB is no more consistent with 
the cross--calibration uncertainties for the instruments, with values 
that differ of more than 25\%. 
The $\chi^2$ associated with this option is significantly higher than 
the broken power-law option, for all the observations. 
We performed a second test by fixing the absorption column density 
at the Galactic value, in the case of a single power--law. 
The inter--calibration factor in this case is even less consistent with 
the acceptable cross--calibration uncertainties (i.e.\ $>35-40\%$), 
along with higher $\chi^2$ values. 
The results from these tests confirm the existence of an intrinsic spectral curvature 
within the observed 0.3--79 keV energy band.

\section{Black hole Mass Estimate}
\label{sec-mass}

\begin{figure*}
\vskip -1 true cm 
\hskip -0.5cm
\includegraphics[width=0.5\textwidth]{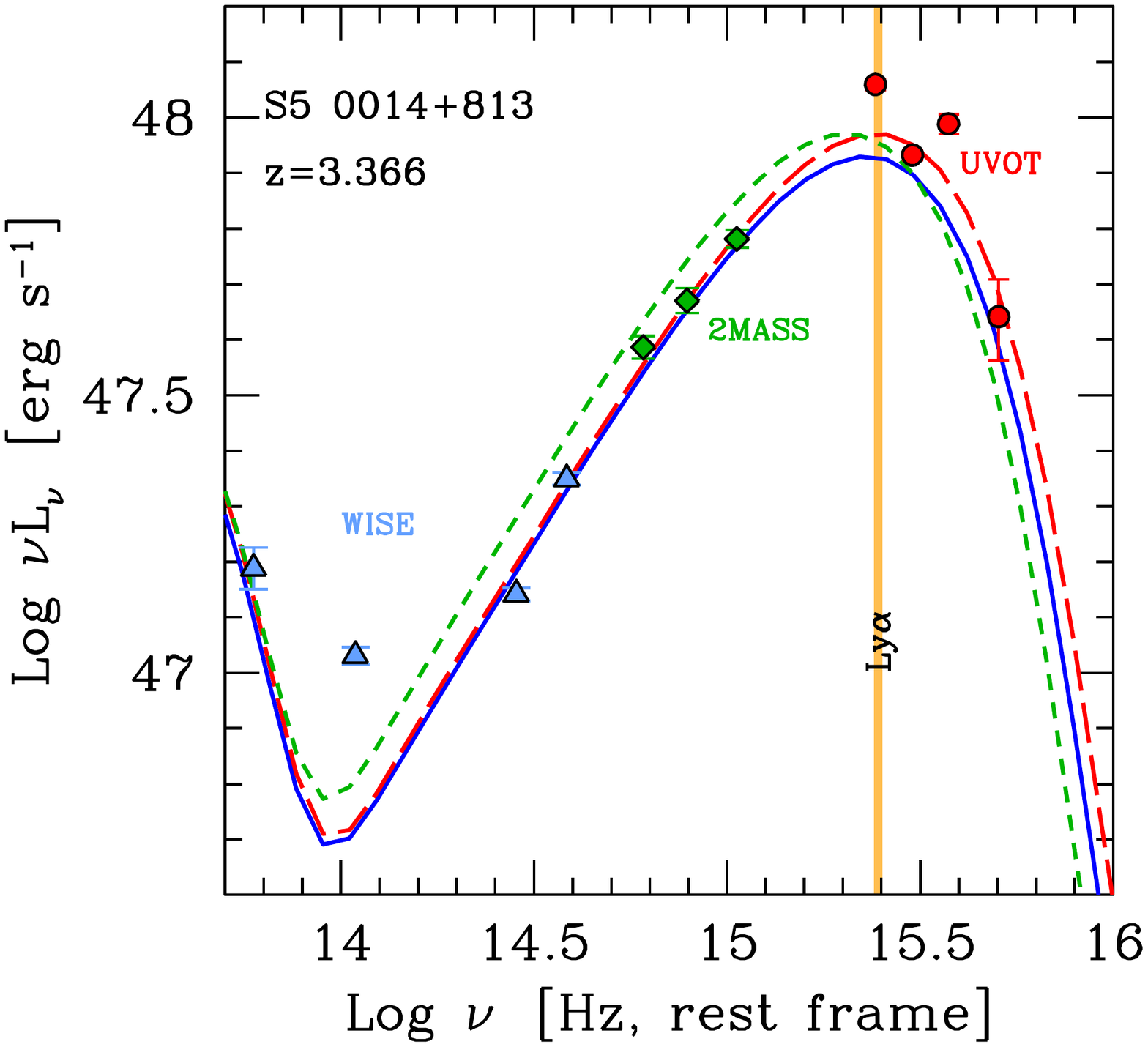}
\includegraphics[width=0.5\textwidth]{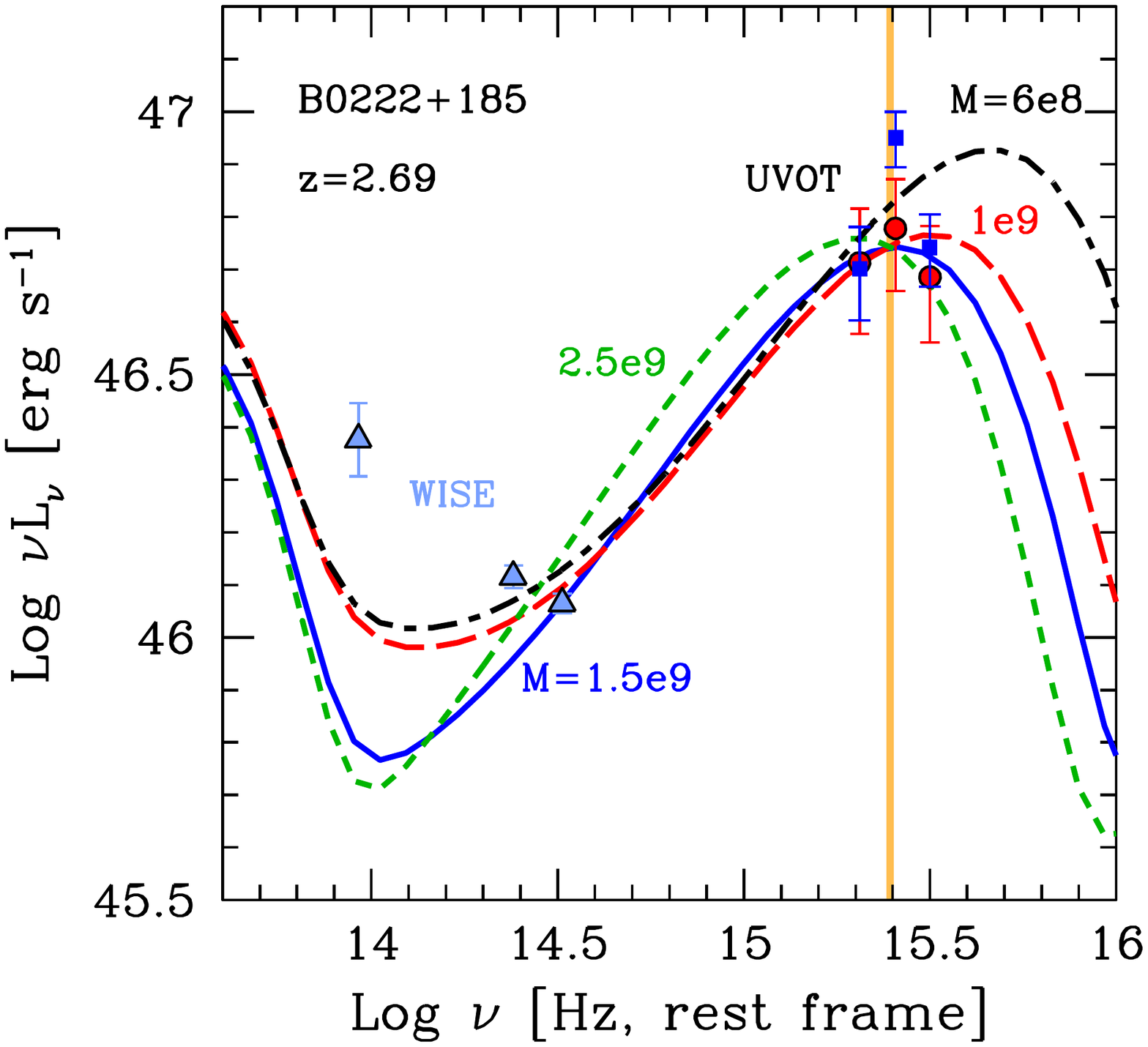}
\vskip -2.5 true cm
\caption{
IR--optical--UV SEDs of S5 0014+813 and B0222+185, along with models 
used to derive $L_{\rm d}$ and $M_{\rm BH}$.
In both panels, the yellow vertical line highlight the Ly$\alpha$ line position. 
At frequencies larger than this line, the {\it Swift}/UVOT data points 
have been corrected for the absorption by intervening clouds. 
The cloud distribution, though, is random and strongly dependent on the line--of--sight. 
Hence the absorption correction is only statistical, with little 
reliability on single sources. 
We do not consider this wavelength range for our modeling. 
{\it Left panel}: S5 0014+815 IR--optical--UV SED. 
The red circles are new {\it Swift}/UVOT data. 
Green diamonds are 2MASS and blue triangles are {\it WISE} data points. 
The solid blue line 
corresponds to our best model, with 
$M_{\rm BH}=7.5\times10^9M_\odot$ and 
$L_{\rm d}=0.85L_{\rm Edd}=8.3\times10^{47}\ergs$.
The green short--dashed line shows a model with 
$M_{\rm BH}=10^{10}M_\odot$ and 
$L_{\rm d}=0.7L_{\rm Edd}=9.1\times10^{47}\ergs$, 
while the red long--dashed line with 
$M_{\rm BH}=7\times10^9M_\odot$ and 
$L_{\rm d}=L_{\rm Edd}=9.1\times10^{47}\ergs$.
The last two represent the ``confidence range'' of $M_{\rm BH}$ 
and $L_{\rm d}$.   
{\it Right panel}: B0222+183 IR--optical--UV SED. 
Red circles are {\it Swift}/UVOT data from the January observation, 
blue squares from the December observation. 
Blue triangles are archival {\it WISE} data points. 
The solid blue line is the best model, with 
$M_{\rm BH}=1.5\times10^9M_\odot$ and 
$L_{\rm d}=0.27L_{\rm Edd}=5.3\times10^{46}\ergs$. 
The red long--dashed line corresponds to
$M_{\rm BH}=1\times10^9M_\odot$ and 
$L_{\rm d}=0.42L_{\rm Edd}=5.5\times10^{46}\ergs$,  
while the green short--dashed to 
$M_{\rm BH}=2.5\times10^9M_\odot$ and 
$L_{\rm d}=0.17L_{\rm Edd}=5.5\times10^{46}\ergs$. 
The black dot--dashed line represents instead a tentative 
model obtained by trying to fit the UVOT December data 
at $\nu=10^{15.4}$Hz. 
The last two represent the ``confidence range'' of $M_{\rm BH}$ 
and $L_{\rm d}$.   
} 
\label{disk_zoom}
\end{figure*}

The black hole mass of a blazar is an important feature to characterise it. 
When optical or infrared spectra are not available, the virial mass estimate 
method cannot be applied. 
At high redshift, the available lines to apply such a method are also less reliable. 
With a good photometric coverage of the UV--optical--IR band, 
instead, the accretion disk emission can be fitted, and the black hole mass 
can be estimated from this fitting process (Calderone et al.\ 2013). 

Figure \ref{disk_zoom} shows the optical--UV band of the SEDs, 
along with {\it Swift}/UVOT photometric data, archival data from 
the ASI Science Data Center (ASDC\footnote{\url{www.asdc.asi.it}}), 
that cover well the whole band.
First we need to point out that at frequencies higher than $\log(\nu/{\rm Hz})=15.4$ rest frame 
a prominent absorption feature is usually present, i.e.\  the Ly$\alpha$ forest, due
to intervening clouds absorbing hydrogen Ly$\alpha$ photons
at wavelengths shortward of 1216 \AA.
We corrected for the absorption following Ghisellini et al.\ (2010). 
However, the distribution of intervening clouds varies randomly along 
every line--of--sight. 
Only an average correction can be done, and it is not sufficiently reliable 
when applied on single sources. 
For this reason, we do not consider the data point at $\log(\nu/{\rm Hz})>15.4$ 
in our modeling. 
The bluer UVOT bands fall in this frequency range in the case of 
S5 0014+318 and B0222+185. 
At frequencies lower than this prominent absorption feature, though, 
a peak in the SEDs is clearly visible.
Below this peak, the optical flux decreases with frequency, suggesting 
a power--law trend, especially in S5 0014+813. 
This is the clear signature of an accretion disk, which can be fitted with a simple 
Shakura--Sunyaev model (Shakura \& Sunyaev 1973; Calderone et al.\ 2013; 
Sbarrato et al.\ 2013). 

At lower frequencies, another feature is evident from the IR--optical--UV SEDs 
of these two sources: the {\it WISE} IR bands show an increase of the flux, 
that breaks the power--law--like trend in the optical. 
This is likely the signature of the IR emission from a dusty torus around the nucleus. 
Such a steep IR spectrum, in fact, could not be produced by synchrotron emission: 
either a self--absorption frequency larger that $\sim10^{13}$Hz or a steep thin synchrotron 
spectrum ending with an exponential cut would be needed to justify such a spectral profile. 
Both options would show up with prominent signatures in the high--energy 
emission of the sources, that we do not observe. 
Figure \ref{disk_zoom} shows that our models do not perfectly reproduce far--IR data. 
In fact our model over--simplifies the torus emission: we describe it as a black body emission, 
while there is evidence (see e.g.\ Calderone, Sbarrato \& Ghisellini 2012) that it 
is best represented as a multi--temperature structure. 
The hottest part, closer to the accretion disk, has likely a temperature $\sim2000$K, 
i.e.\ of the order of the dust sublimation temperature.  
Our far--IR data, in fact, show an increase in flux at frequencies $\sim1.6\times10^{14}$Hz, 
that roughly corresponds to these range of temperatures. 

With these premises, a reliable way to estimate the black hole mass of 
S5 0014+318 and B0222+185 is to fit their IR--optical--UV SEDs 
with a simple model of accretion disk emission. 
We applied the radiatively efficient, geometrically thin, optically thick Shakura \& Sunyaev (1973) model. 
Assuming a standard radiative efficiency $\eta=0.08$, only two free 
parameters are left to be fitted: the accretion rate $\dot M$, that can be traced 
by the intrinsic disk luminosity $L_{\rm d}=\eta\dot Mc^2$, and the black hole mass 
$M_{\rm BH}$ itself. 
In the case of S5 0014+318, 
we can constrain the overall disk luminosity thanks to the visibility of the peak of the disk emission, 
with some consideration regarding its anisotropic properties (as thoroughly explained by Calderone et al.\ 2013):
\begin{enumerate}
\item
according to the Shakura--Sunyaev model, the peak luminosity 
$\nu_p L_{\nu_p}$ corresponds to half the total observed luminosity 
$L_{\rm obs}=2\nu_p L_{\nu_p}$. 

\item 
the observed luminosity depends on the viewing angle of the accretion disk:
\begin{equation}
L_{\rm obs} = 2\cos\theta_{\rm v}L_{\rm d}
\end{equation}
where $L_{\rm d}$ is the intrinsic total luminosity emitted by the accretion disk. 
In the case of a blazar $L_{\rm obs}\simeq2L_{\rm d}$ 
since we see the accretion disk face--on.  
\end{enumerate}
We can therefore derive the intrinsic total luminosity from the peak luminosity 
of our sources:
\begin{equation}
L_{\rm d} = \frac{\nu_p L_{\nu_p}}{\cos\theta_{\rm v}} \simeq \nu_p L_{\nu_p}.
\end{equation}
This means that in the case of S5 0014+318, $L_{\rm d}$ is constrained by 
observations (i.e.\ 2MASS and UVOT data), 
and only $M_{\rm BH}$ is left as a free parameter to be derived 
with the IR--optical--UV SED fitting. 

We find that both sources have large black hole masses and are fast 
accreting, even if not super--Eddington. 
We derive $M_{\rm BH}=1.5\times 10^9 M_\odot$ and $L_{\rm d}=5.3\times 10^{46}$ $\ergs$ 
for B0222+185 and 
$M_{\rm BH}=7.5\times 10^9 M_\odot$ and $L_{\rm d}=8.3\times 10^{47}$erg s$^{-1}$ for S5 0014+813.
These values are significantly smaller than what was derived in Ghisellini et al. (2009 
and 2010a). 
For S5 0014+813 the reason is due to i) the better coverage of the IR band achieved
with {\it WISE} data and ii) neglecting the optical data taken from Bechtold et al. (1994).
We now prefer to discard those data because the derivation of flux and luminosities are
not sufficiently clear in that paper.
The confidence range of S5 0014+813 mass ($7\times10^9-10^{10}M_\odot$) 
is indicated by the dashed lines in the left panel of Fig.\ \ref{disk_zoom}. 
Note that it strongly depends on data quality. 
In this case, the range is rather narrow because of very good data. 
More precise data would lead to even more refined estimates. 
A lower limit on the mass is anyway fixed by the strong constraint given by the Eddington limit. 

For B0222+185, the smaller values of $M_{\rm BH}$ and $L_{\rm d}$ are due to the new infrared data 
(not available in the previous work), that now help 
in roughly constraining the peak frequency of the disk emission. 
The data coverage in this case is not enough to constrain the peak frequency 
luminosity, therefore the estimate on the black hole mass is less constrained. 
This should be taken as an indication of $M_{\rm BH}$, not as a best fit. 

A side result of $M_{\rm BH}$ and $L_{\rm d}$ studies is an estimate of 
the broad line region covering factor with respect to the accretion disc $f_{\rm BLR}$. 
The BLR is thought to reprocess a fraction $f_{\rm BLR}\sim0.05-0.2$ of the 
radiation emitted from the disk, thus usually a standard value $f_{\rm BLR}\simeq0.1$
is used. 
When BLR and disk luminosities are obtained independently, 
the BLR covering factor can be derived, and this is the case. 
Cao \& Jiang (1999) derived $L_{\rm BLR}=4.348\times10^{46}\ergs$ 
for S5 0014+813.
By comparing it with our result, we obtain a covering factor $f_{\rm BLR}=0.05$.

\begin{figure*}
\vskip -1 true cm 
\hskip -0.5cm
\includegraphics[width=0.5\textwidth]{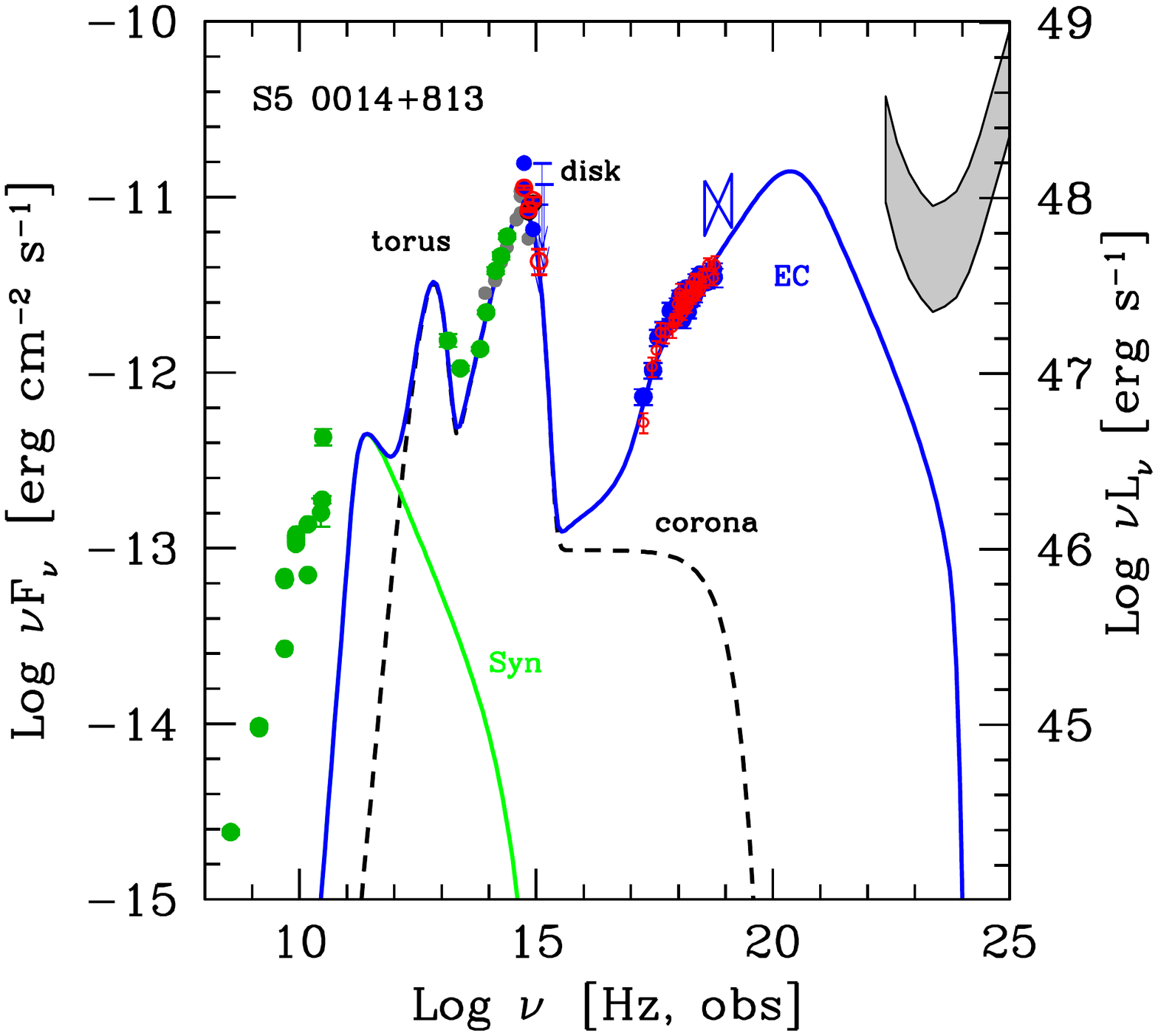}
\includegraphics[width=0.5\textwidth]{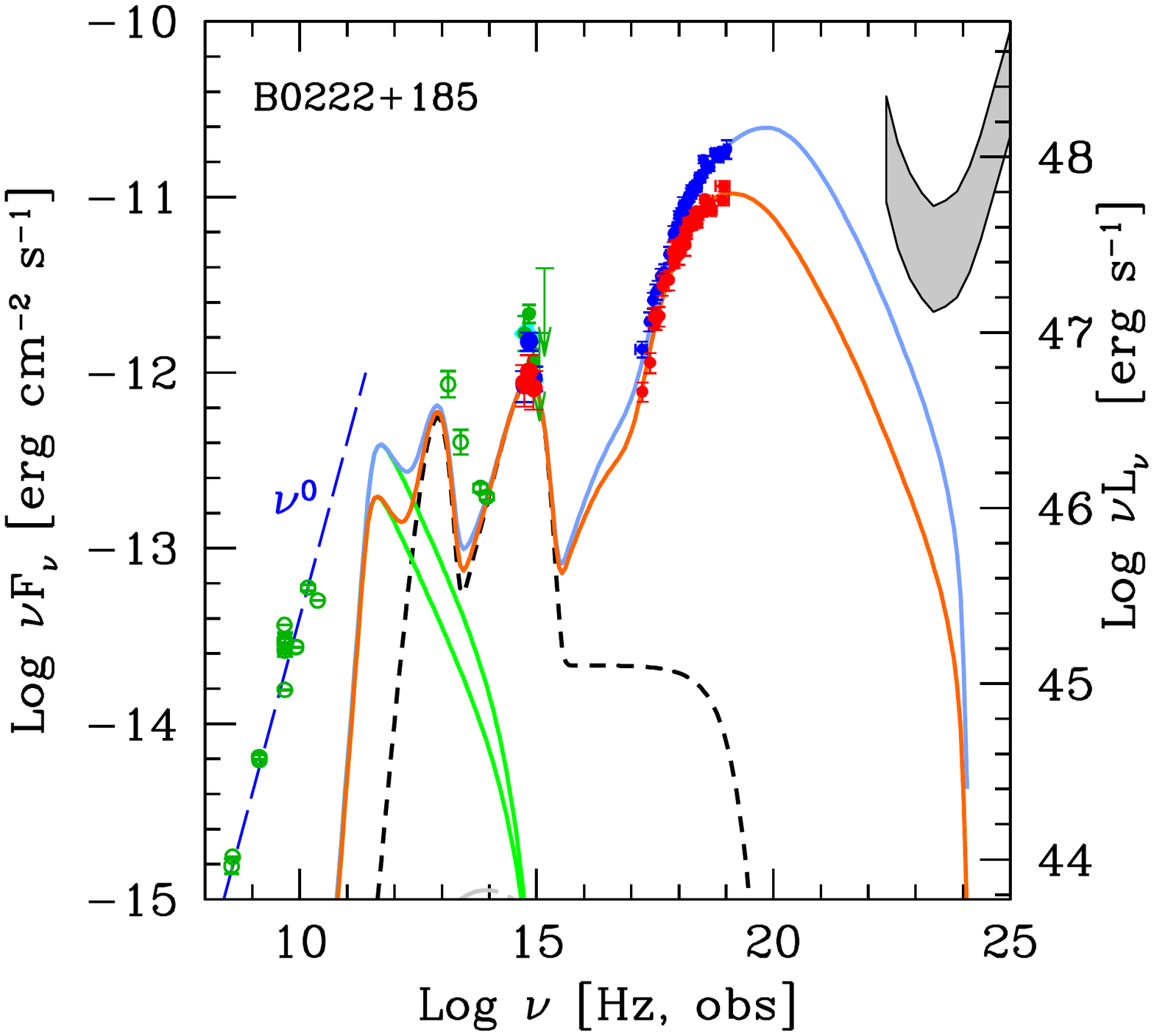}
\vskip -2.5 true cm
\caption{
Broad--band SEDs of S5 0014+813 and B0222+185 along with the models 
discussed in the text and parametrised as in Tables \ref{para} and \ref{powers}. 
In both panels, the grey stripe is the $5\sigma$ {\it Fermi}/LAT sensitivity limit, 
calculated for 5 years (lower edge) and 1 year of operations (upper edge). 
New {\it Swift}/UVOT, {\it Swift}/XRT and {\it NuSTAR} data for the two observation periods 
are red (January 2015) and blue (December 2014) circles. 
{\it Left panel}: S5 0014+813 SED with its broad--band model (blue solid line). 
The green solid line is the self--absorbed synchrotron emission, while the thermal 
emission from accretion disk, torus and X--ray corona is shown with the dashed black line. 
Green data points in radio and IR are from the literature (for details see Ghisellini et al.\ 2009). 
Archival {\it Swift}/BAT data are shown in blue.
{\it Right panel}: SED of B0222+185 with the two models corresponding to the low state 
(orange solid line) and the high state (blue solid line). 
Synchrotron emission in the two states are both shown with green solid lines. 
The thermal emission from the nuclear region (black dashed line) does not 
vary between the two states. 
Green empty circles are archival data (from ASDC). 
} 
\label{sed1}
\end{figure*}

\begin{table*}
\centering
\footnotesize
\begin{tabular}{lllll lllll lllll ll}
\hline
\hline
Source &$z$ &$M_{\rm BH}$ &$R_{\rm diss}$ &$R_{\rm BLR}$ &$R_{\rm T}$  &$L_{\rm d}$ &$L_{\rm d}/L_{\rm Edd}$ &$\Gamma$ &$\theta_{\rm v}$    
&$P^\prime_{\rm i}$ &$B$ &$\gamma_{\rm b}$ &$\gamma_{\rm max}$ &$\gamma_{\rm cool}$ &$s_1$  &$s_2$ \\
~[1]       &[2] &[3] &[4] &[5] &[6] &[7] &[8] &[9] &[10] &[11] &[12] &[13] &[14] &[15] &[16] &[17] \\ 
\hline
S5 0014+813   &3.336 &7.5e9 &1350 &2878 &7.2e4 &829  &0.85 &14 &3  &0.042  &3.5  &45 &2e3 &1   &--1 &3.3  \\
B0222+185 H &2.690  &1.5e9 &360  &725  &1.8e4 &52.7 &0.27 &14 &3  &0.065  &3.3  &35 &2e3 &3.2 &1   &3.2  \\
B0222+185 L &2.690  &1.5e9 &360  &725  &1.8e4 &52.7 &0.27 &14 &3  &0.032  &4.7  &17 &2e3 &3   &1.2 &3.2  \\
\hline
\hline
\end{tabular}
\vskip 0.4 true cm
\caption{Input parameters used to model the SED.
Col. [1]: Source name. H indicates the higher state, L the lower; 
Col. [2]: redshift;
Col. [3]: Black hole mass in solar mass units 
	     (see Figure \ref{disk_zoom} and \S\ref{sec-mass} for the confidence range);
Col. [4]: distance of the blob from the black hole in units of $10^{15}$ cm. 
	     The size of the emitting region is defined as $R_{\rm blob}=\psi R_{\rm diss}$, 
	     where $\psi=0.1{\rm rad}$ is the jet aperture angle;
Col. [5]: radius of the BLR in units of $10^{15}$ cm;
Col. [6]: radius of the torus in units of $10^{15}$ cm;
Col. [7]: disk luminosity in units of $10^{45}$ erg s$^{-1}$ 
	     (see Figure \ref{disk_zoom} and \S\ref{sec-mass} for the confidence range);
Col. [8]: disk luminosity in units of the Eddington luminosity;
Col. [9]: bulk Lorentz factor;
Col. [10]: viewing angle (degrees);
Col. [11]: power injected in the blob calculated in the comoving frame, in units of $10^{45}$ erg s$^{-1}$;
Col. [12]: magnetic field in Gauss;
Col. [13], [14]: break and maximum random Lorentz factors of the injected electrons;
Col. [15]: random Lorentz factors of the electrons cooling in $R/c$;
Col. [16] and [17]: slopes of the injected electron distribution [$Q(\gamma)$] below
 and above $\gamma_{\rm b}$.
The spectral shape of the corona is assumed to be $\propto \nu^{-1} \exp(-h\nu/150~{\rm keV})$.
}
\label{para}
\end{table*}

\begin{table*}
\centering
\begin{tabular}{lllll }
\hline
\hline
Source &$\log P_{\rm r}$ &$\log P_{\rm B}$ &$\log P_{\rm e}$ &$\log P_{\rm p}$ \\
~[1]       &[2] &[3] &[4] &[5]  \\
\hline
S5 0014+813   &46.4 &47.2 &44.6 &47.3 \\
B0222+185 H &46.5 &46.0 &45.5 &48.2 \\
B0222+185 L &46.1 &46.3 &45.3 &48.1 \\
\hline
\hline
\end{tabular}
\vskip 0.4 true cm
\caption{Logarithm of the jet power in different forms. 
Col. [1]: Source name. H indicates the higher state, L the lower; 
Col. [2]: jet power in the form of radiation;
Col. [3]: jet power connected to Poynting flux; 
Col. [4]: jet power in form of bulk motion of electrons; 
Col. [5]: jet power in form of bulk motion of protons (assuming one cold proton per emitting electron). 
}
\label{powers}
\end{table*}


\section{Modelling the broad--band SED}
\label{sec-sed}

Figure \ref{sed1} shows that both S5 0014+318 and B0222+185 have 
overall SEDs characterised by a prominent high--energy component, 
that along with the characteristic flat and intense radio luminosity is 
attributed to non--thermal emission from a relativistic jet. 
In the IR--optical--UV range of both sources, the SEDs are dominated by 
thermal emission attributed to the accretion disk, as discussed in Section \ref{sec-mass}. 

Not being detected in the $\gamma$--rays by {\it Fermi}/LAT, X--ray data 
are necessary to study the non--thermal high--energy emission of 
S5 0014+318 and B0222+185. 
Specifically, {\it NuSTAR} data are crucial for understanding the X--ray spectral profile 
and possible variability in this kind of high--redshift source, as can be seen 
in Figure \ref{zoom}. 
X--ray data contribute significantly to the modelling of 
the broad--band SEDs of the two sources (Figure \ref{sed1}). 

To interpret the SEDs of the two sources, we used a leptonic one--zone 
emitting model, fully described in Ghisellini \& Tavecchio (2009). 
We refer to the original paper for details, providing here only a very brief 
description of the most important features of the models. 
The emitting source is assumed to be a spherical region in which relativistic 
electrons emit by synchrotron and inverse Compton processes. 
This homogeneous spherical blob is assumed to be located at a distance $R_{\rm diss}$ from 
the central black hole, moving with a bulk Lorentz factor $\Gamma$ 
at an angle $\theta_{\rm v}$ from our line--of--sight. 
Relativistic electrons are injected throughout the source, with a power $P^\prime_{\rm i}$ as
measured in the comoving frame. The energy distribution $Q(\gamma)$ of the injected electrons is a 
smoothly broken power law with slopes $s_1$ and $s_2$ (defined as $Q(\gamma)\propto \gamma^{-s}$) 
below and above the random Lorentz factor $\gamma_{\rm b}$.
Note that, even if $Q(\gamma)$ is a broken power law,  
the particle energy distribution $N_\gamma$ derived through the continuity equation
maintains a break, albeit smoother than the injected broken power law. 
This produces a gently curved spectrum, as shown in Figure 1. 
The broad line region is located at a distance $R_{\rm BLR}=10^{17}L_{\rm d, 45}^{1/2}{\rm cm}$ 
from the black hole, while the infrared emitting torus is at 
$R_{\rm torus}=2.5\times10^{18}L_{\rm d, 45}^{1/2}{\rm cm}$. 
$L_{\rm d, 45}$ is the accretion disk luminosity in units of $10^{45}\ergs$, and 
it is derived as in Section \ref{sec-mass}, together with the central black hole mass. 
The values of the parameters adopted for the models are reported in Table \ref{para}.
Note that the model we apply is very sensitive to changes in the derived parameters. 
The emission profile and intensity reproduced by the SED fitting change significantly 
even after small parameter variations, as shown for small variations in the viewing angle
in Fig.\ 3 of Sbarrato et al.\ (2015). 

Table \ref{powers} reports the different forms of the power carried by the jet:
the power $P_{\rm r}$ spent in producing the radiation we observe, 
the Poynting flux $P_{\rm B}$, 
the power associated to the bulk motion of relativistic electrons 
($P_{\rm e}$) and cold protons ($P_{\rm p}$), assuming one proton per relativistic electron. 
This assumption is consistent with independent results on blazar and GRB jets by Nemmen et al.\ (2012). 
They found that the total jet power for both classes is ten times the radiative power $P_{\rm rad}$, 
i.e.\ similar to what we find in this work (see Table \ref{powers}). 
Different proton--to--relativistic electron ratios were explored by Sikora \& Madejski (2000), 
who found that the relativistic pairs must be less than 10-20 per proton. 
With this combination, the total jet power can result equal to or even less than the radiative power, 
that instead is only a part of the total power carried by the jet, and hence 
should be a lower limit to the total $P_{\rm j}$ (Ghisellini 2012; Ghisellini et al.\ 2014). 
Therefore, assuming one proton per relativistic electron is reasonable to explain the 
observed jet features and its physics. 

The emitting regions of both sources are located within the broad line region (and the infrared torus). 
In this way, the energy density of photons from the broad line region feed the 
inverse Compton process, together with photons from the torus. 
The inverse Compton process is dominated by external Compton 
instead of synchrotron--self Compton, as expected in FSRQs. 
Tagliaferri et al.\ (2015) obtained the same result for two other $z>2$ blazars observed 
by {\it NuSTAR}: the emitting regions of both S5 0836+710 and PKS 2149--306 are 
located between the BLR and IR torus. 
Their results were obtained through SED fitting, and were also confirmed on the basis 
of the variability timescales obtained with two {\it NuSTAR} observations per source.

\section{Discussion}
\label{sec-discussion}

Blazars are characterised by their prominent relativistically boosted jet emission. 
They usually show prominent high--energy emission, which results in high 
$\gamma$--ray luminosities, well detected by instruments like {\it Fermi}/LAT. 
In some cases, though, blazars are not detected in such energy bands. 
This is the case for S5 0014+318 and B0222+185\footnote{
Tavecchio et al.\ (2007) predicted with a previous modeling that 
B0222+185 would have been detected by {\it Fermi}/LAT in its 
first year of operation, but this did not happen. 
The authors did not have any IR data, though, to constrain the torus emission 
and hence the synchrotron component. 
Their analysis thus lead to a flatter synchrotron and EC peak, in principle detectable 
by {\it Fermi}/LAT. The steeper slope we now observe thanks to IR data, 
instead, is consistent with non--detection in the $\gamma$--rays.
}. 
Even if lacking a high--energy detection, the Compton 
bump can be observed in the X--ray frequency range, 
but a detection in the soft X--rays usually is not enough to determine 
the relativistic jet features of a blazar, nor its orientation.  
S5 0014+318 and B0222+185 were detected by {\it Swift}/BAT, 
but these data were not precise enough to derive exact estimates of 
the bulk Lorentz factors and viewing angles. 
Figures \ref{zoom} and \ref{sed1} show that the {\it Swift}/BAT data do not 
have enough precision to constrain the hard X--ray slope. 
{\it NuSTAR}, however, provides a broad--band, precise measurement for both sources,  
confirming that both blazars are seen at small viewing angles, i.e.\ 
their jets are directed along our line of sight. 
Both sources also host massive central black holes (both with $M_{\rm BH}>10^9M_\odot$). 

\subsection{Variability}
\label{sec-var}

Looking in detail at the broad--band SEDs of these two objects, we see 
some interesting differences. 
Both sources show the two humps, i.e.\ the signature of aligned jet emission, while 
in the IR--optical--UV band, the accretion disk emission dominates 
over the non--thermal jet emission. 
Comparing the two panels of Figure \ref{sed1}, it can be noticed that 
S5 0014+318 and B0222+185 also show different variability behaviours. 
They were both observed in two epochs separated by $\sim1$ month. 
S5 0014+318 does not show flux or spectral variation between the 
two observations, and the new data are consistent with archival data. 
Only the {\it Swift}/BAT detection could suggest a different state of the source, 
but due to the large uncertainty we cannot draw any strong conclusion. 

B0222+185, instead, shows a clear variation of the X--ray flux between the 
two {\it Swift}/XRT + {\it NuSTAR} observations: in December 2014 the source was in a higher 
state, compared to both January 2015 and archival data. 
The different hard X--ray spectra showed in the two B0222+185 observations 
(see right panel of Figure \ref{zoom}) suggest that the peak of the high--energy hump 
is at the same or higher frequency when in the higher state, compared to 
the lower state. 
This would be opposite to the general trend displayed by the blazar sequence, 
but missing a higher--frequency detection, this speculation is not conclusive. 
If real, such behaviour would not be uncommon in rapidly varying FSRQs: 
although they follow the
blazar sequence when considering different sources,
an individual object can behave opposite to the sequence itself while varying. 
According to the model shown in Figure \ref{sed1}, B0222+185 variability can be 
described by a variation in the injected power (see second and third lines of Table \ref{para}),
accompanied by $\gamma_{\rm b}$ increasing in the high state. 

Another remarkable example of this kind of variation, very similar to the one showed by 
B0222+185, but much more pronounced, has recently been seen for S5 0836+710 
(Ciprini et al.\ 2015; Vercellone et al.\ 2015; Giroletti et al.\ 2015)
during its August 2015 $\gamma$--ray flare, which triggered observations 
at X--ray and radio frequencies. 
The amplitude of the flux variation was huge in the {\it Fermi}/LAT band
(factor 65 greater than the average flux reported in the third 
{\it Fermi}/LAT catalog of Acero et al.\ 2015)
and rather modest in the high energy part of the {\it Swift}/XRT band.
This implies that the X--ray flux had to change more at larger energies,
to connect to the enhanced $\gamma$--ray flux, and that the peak frequency of
the high energy hump must be ``bluer" than what was displayed during the 
{\it NuSTAR} observation described in Tagliaferri et al.\ (2015).
In other words: if {\it NuSTAR} had followed the August 2015 flare of S5 0836+710, 
it likely would have 
detected a clear flux and spectral variation, leading to a predicted 
shift of the high energy hump towards higher frequencies even in the absence of 
$\gamma$--ray data.
The right panel of Figure \ref{zoom} shows that {\it Swift}/XRT would 
not have been able to discriminate between the two states of B0222+185, 
while {\it NuSTAR} distinguishes them clearly.
We conclude that: 
\begin{itemize}
\item[--]
observations in the hard X--ray band are not only instrumental to discover
the most powerful blazars (that requires a survey of a large portion of the sky),
but also to detect large flux variations that occur around the MeV energy band
and are not noticeable in the classic 0.3--10 keV band, nor at energies beyond 
100 MeV, where the source could go undetected even during a large flare; and
\item[--]
repeated hard X--rays observations on timescales of a few weeks up to 1--2 months 
are efficient to spot variability in high--redshift blazars. 
We managed to see variability in B0222+185, with similar features as 
the already observed variable FSRQs S5 0836+710 and PKS 2149--306 
discussed by Tagliaferri et al.\ (2015).
\end{itemize}

\subsection{Jets and accretion of the two most powerful blazars}
\label{sec-jetaccr}

\begin{figure*}
\centering
\vskip -1 cm
\includegraphics[width=10cm]{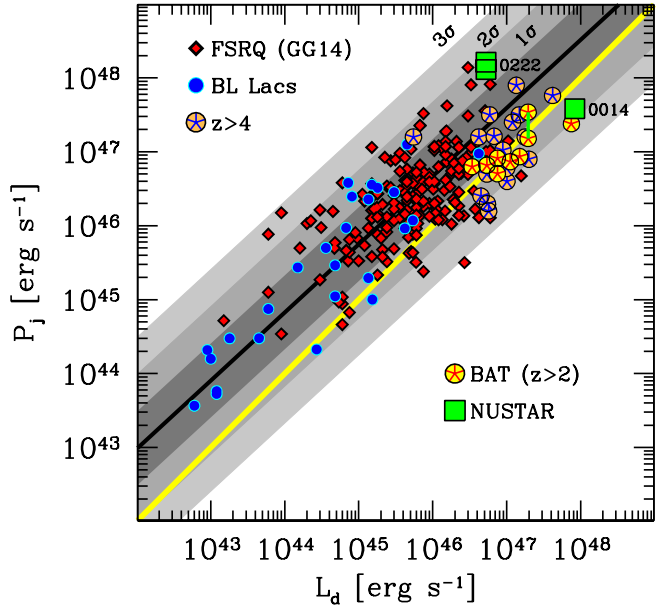}
\vskip -2.5 true cm
\caption{
Total jet power as a function of accretion disk luminosity in blazars. 
Red (outlined in black) diamonds and blue filled circles are respectively 
FSRQs and BL Lacs from Ghisellini et al.\ (2014). 
Circled asterisks are high--redshift, non $\gamma$--ray detected blazars: 
pink--filled circles with blue asterisks are all the known $z>4$ blazars (see Ghisellini et al.\ 2015), 
yellow--filled circles with red asterisks are the $z>2$ blazars detected by {\it Swift}/BAT. 
Green filled squares are S5 0014+318 and B0222+185 (both states) as labelled, 
observed by {\it NuSTAR} for this work. 
Note that the latter are the two blazars with the most luminous accretion disk (S5 0014+318) 
and the most powerful jet (B0222+185). 
The black line and grey stripes are the best fit relation of $\gamma$--ray detected blazars 
from Ghisellini et al.\ (2014), along with the $1\sigma$, $2\sigma$ and $3\sigma$ dispersions. 
The yellow line is the one--to--one correlation. 
S5 0836+710 underwent a prominent $\gamma$--ray flare, and therefore it has been plotted 
with the jet powers of both its average and flaring states. 
The two states are connected with a green line. 
This suggests that part of the spread could be due to different states of the single sources, 
along with flaring episodes. 
The jet power is calculated as the sum of the different components listed in Table \ref{powers}. 
Note that the very powerful blazars we added to the original sample by Ghisellini et al.\ (2014) 
are located within $2\sigma$ from the $\gamma$--ray detected jet--accretion relation. 
This means that also in the most powerful sources the jet power correlates with accretion luminosity, 
but it is larger than the accretion power, leading to an important role of black hole spin in jet launching. 
} 
\label{jet_accr}
\end{figure*}

We now aim to frame S5 0014+318 and B0222+185 within the larger blazar picture. 
We consider them in the jet--accretion correlation scenario. 
Ghisellini et al.\ (2014) found that in blazars the jet power not only correlates 
with the accretion power, but it is even larger. 
This suggests that accretion is strongly related to jet power, implying a role 
in jet production. 
At the same time, the fact that jet power is larger than accretion power 
tells us that some other process must play a role in the jet launch 
and acceleration. 
Black hole spin is the best candidate to play such a role. 
This result was obtained by studying a sample of {\it Fermi}--detected 
blazars, for which Shaw et al.\ (2012; 2013) obtained optical spectra. 
Ghisellini et al.\ (2014) selected all the objects with broad emission lines, 
in order to have a proxy of accretion luminosity, and compared jet 
and accretion power for the 226 blazars in this sample. 
However, this sample did not include the most extreme blazars known, 
leaving open the questions:  
{\it 
how does the jet--accretion relation look in the case of the most powerful blazars? 
}
Does the power balance change when accretion or jet emission are extreme?
These questions will guide us in the following discussion. 

First we add to the original blazar sample the sources expected to be the most powerful. 
To this aim we select the $z>2$ blazars detected by {\it Swift}/BAT and all 
known high--redshift ($z>4$) blazars. 
The BAT sensitivity limit is not very deep, and at high redshift it can detect mainly 
the most powerful sources, whose high--energy components peak in the $\sim$MeV range. 
BAT detected 10 $z>2$ blazars, including S5 0014+318 and B0222+185, that we 
add to the blazar sample of Ghisellini et al.\ (2014). 
We also include all the known blazars at $z>4$, as listed in Ghisellini et al.\ (2015). 
Being the highest redshift blazars currently known, they are expected to be 
among the most powerful blazars.
They are not present in the BAT blazar catalog because their distance makes their 
hard X--ray flux too weak for a detection with BAT.
Since most of them were selected starting from optical catalogs, 
they are likely very powerful in accretion luminosity. 

Figure \ref{jet_accr} shows how these samples are located in the overall jet--accretion 
relation, along with S5 0014+318 and B0222+185. 
The total jet power (calculated as the sum of different jet power components  
listed in Table \ref{powers}) is plotted as a function of the disk luminosity. 
The grey stripes show the best fit of the sample by Ghisellini et al.\ (2014). 
Note that the powerful blazars we add in this work are all located within the 2$\sigma$ 
dispersion of the previous correlation. 
This means that they still follow the jet--accretion relation found by Ghisellini et al.\ (2014), 
even if they are among the most powerful sources in the set. 

S5 0014+318 and B0222+185 can be considered the two most extreme sources: 
respectively, they are the blazars with the most luminous disk and the most powerful jet. 
Still, they are close enough to the known jet--accretion correlation, to be less than 2$\sigma$ 
from the Ghisellini et al.\ (2014) result. 
Thus we conclude that even the most powerful blazars follow the same jet--accretion 
relation as the $\gamma$--ray detected bulk sample. 
The second interesting conclusion that we can draw from this comparison is that 
{\it NuSTAR} is once again confirmed to be the most suitable telescope to study the most 
powerful blazars in our Universe.

\section{Conclusions}
\label{sec-concl}

The simultaneous X--ray observations of S5 0014+318 and B0222+185  
performed with {\it Swift}/XRT and {\it NuSTAR} gave us an interesting insight 
into the jet emissions of these two sources. 
We confirmed their blazar nature, with a refined estimate of their 
bulk Lorentz factors and viewing angles, supported by more precise sets of 
parameters (Tables \ref{para} and \ref{powers}). 
The accretion disk fitting to a more complete data set gave us the possibility to refine 
our previous estimates of  
the black hole mass and accretion luminosity of these two sources,  
implying fast accreting objects with extreme masses of $>10^9M_\odot$. 

The overall SED modelling allowed us to estimate the jet power and 
accretion disk luminosity, allowing a comparison of these two sources with the overall blazar 
jet--accretion relation. 
The two sources are among the most powerful blazars known, and they populate the 
highest disk luminosity and jet power part of the 
jet--accretion correlation (Ghisellini et al.\ 2014; Figure \ref{jet_accr}). 
It is remarkable that a sample formed by the most powerful blazars known is 
still within $2\sigma$ from the correlation derived from a sample 
of blazars whose $\gamma$--ray flux was averaged over two years. 
The mechanisms governing the jet formation and evolution in the most extreme 
sources must not be different from the processes powering the more moderate objects. 
S5 0014+318 and B0222+185 themselves are consistent with the relation 
derived from the $\gamma$--ray blazars, even if they are the blazars 
with the most luminous accretion disk and the most powerful jet, respectively. 

We found a different variability behaviour between the two sources: 
while S5 0014+318 did not vary in the two observation epochs separated by $\sim1$ month, 
B0222+185 shows a clear variation, with an amplitude larger 
at larger frequencies. 
This last feature is the main reason why {\it NuSTAR} is such a crucial 
instrument to study high--redshift and powerful blazars. 
A soft X--ray telescope alone could not see a variability event like the one 
shown by B0222+185: {\it NuSTAR} instead observes at frequencies where 
the amplitude of a flaring activity is large enough to be seen. 
Were we lacking the $\gamma$--ray signature of a flare, {\it NuSTAR} 
could do the job. 

\section*{Acknowledgements}
We thank the referee for her/his comments, that helped us to improve the paper.
We acknowledge financial support from the ASI-INAF grant I/037/12/0.
This work
made use of data from the {\it NuSTAR} mission, a project led by
the California Institute of Technology, managed by the Jet Propulsion
Laboratory, and funded by the National Aeronautics and Space
Administration. We thank the {\it NuSTAR} Operations, Software and
Calibration teams for support with the execution and analysis of
these observations.  
This research has made also use of the {\it NuSTAR} Data Analysis Software
(NuSTARDAS) jointly developed by the ASI Science Data Center (ASDC,
Italy) and the California Institute of Technology (Caltech, USA).
This publication makes use of data products from the {\it Wide--field
Infrared Survey Explorer}, which is a joint project of the University
of California, Los Angeles, and the Jet Propulsion
Laboratory/California Institute of Technology, funded by the National
Aeronautics and Space Administration.
Part of this work is based on archival data, software or on--line services 
provided by the ASI Data Center (ASDC).
This research has made use of the XRT Data Analysis Software (XRTDAS)
developed under the responsibility of the ASI Science Data Center (ASDC), Italy.


\label{lastpage}
\end{document}